\documentstyle[preprint,eqsecnum,aps]{revtex}
\tightenlines                                                                                                                                                                
\newcommand{\be}{\begin{equation}}
\newcommand{\ee}{\end{equation}}
\newcommand{\bea}{\begin{eqnarray}}
\newcommand{\eea}{\end{eqnarray}}
\newcommand{\ba}{\begin{array}}
\newcommand{\ea}{\end{array}}

\newcommand{\ep}{\epsilon}
\newcommand{\Th}{\Theta}
\newcommand{\th}{\theta}

\newcommand{\de}{\delta}

\newcommand{\pa}{\partial}

\newcommand{\no}{\nonumber}

\newcommand{\sres}{\mbox{sres}}
\newcommand{\str}{\mbox{str}}

\begin{document}

\title{Canonical Gauge Equivalences of the \\
sAKNS and sTB Hierarchies}

\author{Jiin-Chang Shaw$^1$ and Ming-Hsien Tu$^2$ }
\address{
$^1$ Department of Applied Mathematics, National Chiao Tung University, \\
Hsinchu, Taiwan, \\
and\\
$^2$ Department of Physics, National Tsing Hua University, \\
Hsinchu, Taiwan
}
\date{\today}
\maketitle

\begin{abstract}
We study the gauge transformations between the supersymmetric
AKNS (sAKNS) and supersymmetric two-boson (sTB) hierarchies.
The Hamiltonian nature of these gauge transformations is
investigated, which turns out to be canonical.
We also obtain the Darboux-B\"acklund transformations for
the sAKNS hierarchy from these gauge transformations.
\end{abstract}

\newpage

\section{Introduction}

During the past ten years, the theory of soliton \cite{FT,Das,Dic} has played an important
role in theoretical and mathematical physics. Especially, the explorations on the
relationship between the integrable models and string theories \cite{Mar}.
On the one hand, several kinds of correlation functions in string theory
are governed by the integrable hierarchy equations (e.g. KdV, KP etc.)\cite{Mar}.
On the other hand, the idea of the supersymmetric extensions of the
integrable systems\cite{MR,K,Math} has motivated people to use them to study the theory of
superstrings\cite{Al}. 

Recently, several  supersymmetric integrable systems have been proposed and
studied (see, for example, \cite{FMR,OP,BD,BKS,AR,AD,DG,P,KST} and references therein). 
In this paper, we discuss only two of them;
the supersymmetric AKNS (sAKNS) hierarchy\cite{AR} and 
the supersymmetric two-boson (sTB) hierarchy\cite{BD}. 
The former was introduced from the study of the reduction
scheme in the constrained KP hierarchy\cite{CKP}, and the latter was constructed
from the supersymmetric extension of the dispersive long water wave equation\cite{Kaup,Broer}.
Both of them have supersymmetric Lax representations, being bi-Hamiltonian,
and have infinite conserved quantities etc. 
Besides of these properties, these two hierarchies can be related to each other
via a gauge transformation\cite{AR}. Sometimes, such transformation
from one hierarchy to the other is called Miura transformation.
However, from our view point, the connection between
these two hierarchies  has not been totally explored. 
The purpose of this work is to provide a deeper understanding about the
gauge transformations between the sAKNS and the sTB hierarchies. 

Our paper is organized as follows: After the
introduction of the Lax formulation of the sAKNS hierarchy in 
Sec. II, we discuss the gauge transformations between the sAKNS and the
sTB hierarchies in Sec. III. Sec. IV is devoted to investigate the canonical
property of these gauge transformations from the bi-Hamiltonian view point.
Our approach follows very closely that of Refs.\cite{Mor,ST1} for other systems.
We then show, in Sec. V, that the Darboux-B\"acklund transformations
(DBTs) for the sAKNS hierarchy itself
can be constructed from these gauge transformations. Concluding remarks
are presented in Sec. VI.

\section{sAKNS hierarchy}
The sAKNS hierarchy \cite{AR} has the Lax operator
of the form
\be
L=\pa+\Phi D^{-1}\Psi
\label{lax}
\ee
which satisfies the hierarchy equations
\be
\frac{\pa L}{\pa t_n}=[L^n_+,L]
\label{laxeq}
\ee
where $D=\pa_{\th}+\th\pa$ is the supercovariant derivative defined on
a $(1|1)$ superspace \cite{D} with coordinates $(x,\th)$. $D^{-1}=\th+\pa_{\th}\pa$ 
is the formal inverse of $D$, which satisfies $D^{-1}D=D^{-1}D=1$.
The coefficients functions $\Phi$ and $\Psi$ are superfields with
proper parity such that $L$ is a bosonic operator. 
It can be proved that (\ref{laxeq}) is consistent with the following equations
\be
\frac{\pa \Phi}{\pa t_n}=(L^n_+\Phi), \qquad 
\frac{\pa \Psi}{\pa t_n}=-((L^n)^*_+\Psi).
\label{wfeq}
\ee
where the conjugate operation ``$\ast$" is defined by $(AB)^*=(-1)^{|A||B|}B^*A^*$
for the super-pseudo-differential operator $A,B$ and $f^*=f$ for arbitrary superfield $f$.
Therefore $\Phi$ and $\Psi$ are eigenfunction and adjoint eigenfunction
of the hierarchy, respectively.

Since the Lax operator (\ref{lax}) is assume to be homogeneous under $Z_2$-grading,
the gradings of the (adjoint) eigenfunction should satisfy $|\Phi|+|\Psi|=1$. Here we
refer the parity of a superfield $f$ to be even if $|f|=0$ and odd if $|f|=1$. 
There are two cases to be discussed.

(a) $|\Phi|=0$ and $|\Psi|=1$

In this case, the parity of the eigenfunction $\Phi$ is even, whereas the adjoint
eigenfunction $\Psi$ is odd. Therefore we can parametrize them as
\be
\Phi=\phi_1+\th\psi_1,\qquad \Psi=\psi_2+\th\phi_2.
\label{paraa}
\ee
Substituting (\ref{paraa}) into (\ref{wfeq}), we obtain, for example, the $t_2$-flow as
\bea
\frac{\pa\phi_1}{\pa t_2}&=& \phi_{1xx}+2\phi_1^2\phi_2+2\phi_1\psi_1\psi_2\no\\
\frac{\pa\psi_1}{\pa t_2}&=& \psi_{1xx}+2\phi_1^2\psi_{2x}+2\phi_1\phi_2\psi_1
+2\phi_1\phi_{1x}\psi_2\no\\
\frac{\pa\phi_2}{\pa t_2}&=& -\phi_{2xx}-2\phi_1\phi_2^2-2\phi_2\psi_1\psi_2
+2\phi_1\psi_2\psi_{2x}\no\\
\frac{\pa\psi_2}{\pa t_2}&=&-\psi_{2xx}-2\phi_1\phi_2\psi_2. 
\label{eqa}
\eea
In the bosonic limit ($\psi_1, \psi_2 \rightarrow 0$) the AKNS equations ($t_2$-flow)
are recovered. In fact, this can be easily viewed from the bosonic limit of the
Lax operator $L_{bosonic}=\pa+\phi_1\pa^{-1}\phi_2$.

(b) $|\Phi|=1$ and $|\Psi|=0$

In this case, we have odd eigenfunction $\Phi$ and even adjoint eigenfunction $\Psi$.
Hence, they should be parametrized as
\be
\Phi=\psi_1+\th\phi_1, \qquad \Psi=\phi_2+\th\psi_2
\label{parab}
\ee
The hierarchy equations (\ref{wfeq})(e.g. $t_2$-flow) then become
\bea
\frac{\pa\phi_1}{\pa t_2}&=& \phi_{1xx}+2\phi_1^2\phi_2-
2\phi_1\psi_1\psi_2-2\phi_2\psi_1\psi_{1x}\no\\
\frac{\pa\psi_1}{\pa t_2}&=& \psi_{1xx}+2\phi_1\phi_2\psi_1\no\\
\frac{\pa\phi_2}{\pa t_2}&=& -\phi_{2xx}-2\phi_1\phi_2^2+2\phi_2\psi_1\psi_2\no\\
\frac{\pa\psi_2}{\pa t_2}&=&-\psi_{2xx}-2\phi_2^2\psi_{1x}-
2\phi_1\phi_2\psi_2-2\phi_2\phi_{2x}\psi_1. 
\label{eqb}
\eea
which also contain the ordinary AKNS equations in the bosonic limit.
However, we want to point out that the Lax operator has no direct bosonic
limit in this case. 

It has been shown \cite{AR} that the hierarchy equations (\ref{laxeq}) 
are invariant under the supersymmetric transformations:
\be
\de_{\ep}\Phi=\ep (D^{\dag}\Phi),\qquad \de_{\ep}\Psi=\ep (D^{\dag}\Psi)
\label{susy}
\ee
where $\ep$ is an odd constant and $D^{\dag}\equiv \pa_{\th}-\th\pa$. 
In particular, using the parametrizations (\ref{paraa}) and 
(\ref{parab}), it is straightforward  to show that (\ref{eqa}) and (\ref{eqb}) are
invariant under the transformations (\ref{susy}). In the following sections, 
the sAKNS Lax operators for the case (a) and case (b) will be denoted by
$L_a=\pa+\Phi_a D^{-1}\Psi_a$ and $L_b=\pa+\Phi_b D^{-1}\Psi_b$, 
respectively and thus $|\Phi_a|=|\Psi_b|=0$ and $|\Psi_a|=|\Phi_b|=1$.

\section{Gauge transformations and sTB hierarchy}

Given a sAKNS hierarchy we can construct a nonstandard Lax hierarchy
via a gauge transformation. For case (a), let us perform the following
transformation
\bea
G_a:\quad L_a\rightarrow K &=& \Phi_a^{-1}L_a\Phi_a 
\label{ga}\\
&\equiv&\pa-(DJ_0)+D^{-1}J_1\no
\eea
where both $J_0$ and $J_1$ are odd superfields which can be expressed in terms of
$\Phi_a$ and $\Psi_a$ as follows
\be
J_0=-(D\ln \Phi_a),\qquad J_1=\Phi_a\Psi_a.
\label{dicga}
\ee
The hierarchy equations then become
\be
\frac{\pa K}{\pa t_n}=[K^n_{\geq 1},K]
\label{keq}
\ee
which is the so-called sTB hierarchy\cite{BD}. 
The first few hierarchy equations are given by
\bea
\frac{\pa J_0}{\pa t_1}&=&J_{0x}\no\\
\frac{\pa J_1}{\pa t_1}&=&J_{1x}\no\\
\frac{\pa J_0}{\pa t_2}&=&J_{oxx}-2J_{1x}-(D(DJ_0)^2)\no\\
\frac{\pa J_1}{\pa t_2}&=&-J_{1xx}-2(J_1(DJ_0))_x
\label{stbeq}
\eea
etc. It is easy to show that (\ref{stbeq}) contain the 
Kaup-Broer hierarchy equations\cite{Kaup,Broer}
in the bosonic limit. Substituting (\ref{dicga}) into (\ref{stbeq}) and 
using the parametrizations (\ref{paraa}), one can recover the equations (\ref{eqa}).
Moreover, the hierarchy equations (\ref{keq}) have been
shown \cite{BD} to be invariant under the supersymmetric transformations:
$\de_{\ep}J_0=\ep(D^{\dag}J_0)$, $\de_{\ep}J_1=\ep(D^{\dag}J_1)$.

For case (b), we need another gauge transformation to do the job since
$|\Phi_b|=1$ in this case. Let us consider the following transformation
\bea
G_b: \quad L_B\rightarrow K&=&D^{-1}\Psi_b L_b\Psi_b^{-1}D
\label{gb}\\
&\equiv&\pa-(DJ_0)+D^{-1}J_1\no
\eea
which implies
\be
J_0=(D\ln\Psi_b),\qquad J_1=\Phi_b\Psi_b+(D^3\ln \Psi_b)
\label{dicgb}
\ee
and the Lax operator $K$ still satisfying the hierarchy equations (\ref{keq}).
Substituting (\ref{dicgb}) into (\ref{stbeq}) and using (\ref{parab}), 
we obtain (\ref{eqb}) immediately.  

In fact, both gauge transformations $G_a$ and $G_b$ have their inverse 
transformations $H_a$ and $H_b$, respectively. 
In other words, for a given sTB Lax operator $K$, 
one can perform the following transformation to gauge away the 
constant term and to obtain the Lax operator $L_a$\cite{AR}
\bea
H_a: K\rightarrow L_a&=&e^{-\int^x(DJ_0)}Ke^{\int^x(DJ_0)}
\label{ha}\\
&\equiv&\pa+\Phi_a D^{-1}\Psi_a\no
\eea
where
\be
\Phi_a=e^{-\int^x(DJ_0)},\qquad \Psi_a=J_1e^{\int^x(DJ_0)}.
\label{dicha}
\ee
It can be proved that $L_a$ satisfies (\ref{laxeq}) if $K$ satisfies (\ref{keq}).

Similarly, for case (b), we have
\bea
H_b: K\rightarrow L_b&=&e^{-\int^x(DJ_0)}DKD^{-1}e^{\int^x(DJ_0)}
\label{hb}\\
&\equiv&\pa+\Phi_b D^{-1}\Psi_b\no
\eea
where
\be
\Phi_b=(J_1-J_{0x})e^{-\int^x(DJ_0)},\qquad \Psi_b=e^{\int^x(DJ_0)}.
\label{dichb}
\ee
Since $H_a$($H_b$) is the inverse of $G_a$($G_b$) and vice versa, thus
we obtain the correspondences between the sAKNS and sTB
hierarchies.

\section{Canonical property and Hamiltonian structures}

The discussions presented in the previous section establish the gauge equivalences
between the sAKNS and the sTB hierarchies at Lax formulation level. 
In this section, we would like to discuss the Hamiltonian nature of 
these gauge transformations. Let us start from the sTB hierarchy.

The Lax equation (\ref{keq}) of the sTB hierarchy has a bi-Hamiltonian description
as follows
\be
\pa_{t_n}
\left(
\ba{c}
J_0\\
J_1
\ea
\right)=
\Th_1
\left(
\ba{c}
\frac{\de H_{n+1}}{\de J_0}\\
\frac{\de H_{n+1}}{\de J_1}
\ea
\right)=
\Th_2
\left(
\ba{c}
\frac{\de H_{n}}{\de J_0}\\
\frac{\de H_{n}}{\de J_1}
\ea
\right)
\ee
where the first structure $\Th_1$ and the second 
structure $\Th_2$ are given by\cite{BD}
\bea
\Th_1&=&
\left(
\ba{cc}
0 & -D\\
-D & 0
\ea
\right)\\
\Th_2&=&
\left(
\ba{cc}
2D+2D^{-1}J_1D^{-1}- D^{-1}J_{0x}D^{-1}& 
-D^3+D(DJ_0)-D^{-1}J_1D\\
D^3+(DJ_0)D+DJ_1D^{-1} & J_1D^2+D^2J_1
\ea
\right).
\eea
which have been investigated \cite{BD} to be compatible by 
using the prolongation method \cite{Olver}. 
The Hamiltonians $H_n$ are defined by
\be
H_n=\frac{-1}{n}\str K^n\equiv\frac{-1}{n}\int dxd\th\sres K^n
\ee
where the $\sres$ picks up the coefficient of the $D^{-1}$ term of a
super-pseudo-differential operator.

Since the bi-Hamiltonian structure is one of the most important properties 
of an integrable system, it is quite natural to ask whether the gauge 
transformations discussed above are canonical or not.  To see this, 
from the gauge transformation $H_a$, we can obtain the linearized map
$H_a'$ and its transposed map $H_a'^{\dag}$ as follows
\be
H_a'=
\left(
\ba{cc}
-\Phi_a D^{-1} & 0\\
\Psi_a D^{-1} & \Phi_a^{-1}
\ea
\right),\qquad
H_a'^{\dag}=
\left(
\ba{cc}
D^{-1}\Phi_a & -D^{-1}\Psi_a\\
0 & \Phi_a^{-1}
\ea
\right)
\label{linha}
\ee
where $\Phi_a$ and $\Psi_a$ are related to $J_0$ and $J_1$ via Eq.(\ref{dicga}) 
(or Eq. (\ref{dicha})). A straightforward calculation shows that
\bea
H_a'\Th_1 H_a'^{\dag} &=&
\left(
\ba{cc}
0 &1\\
-1 & 0
\ea
\right)\equiv P_a \\
H_a'\Th_2 H_a'^{\dag} &=&
\left(
\ba{cc}
-\Phi_a D^{-2}\Phi_a D-D\Phi_a D^{-2}\Phi_a  & D^2+D\Phi_a D^{-2}\Psi_a+\Phi_a
 D^{-2}(D\Psi_a) \\
-2\Phi_a D^{-2}\Phi_a\Psi_a D^{-2}\Phi_a & +
2\Phi_a D^{-2}\Phi_a\Psi_a D^{-2}\Psi_a \\
\mbox{} & \mbox{} \\
D^2+\Psi_a D^{-2}\Phi_a D+(D\Psi_a)D^{-2}\Phi_a &
 -\Psi_a D^{-2}(D\Psi_a)-(D\Psi_a)D^{-2}\Psi_a\\
+2\Psi_a D^{-2}\Phi_a\Psi_a D^{-2}\Phi_a & 
-2\Psi_a D^{-2}\Phi_a\Psi_a D^{-2}\Psi_a
\ea
\right)\equiv Q_a
\eea
where $P_a$ and $Q_a$ are just the first and the second Hamiltonian structures 
obtained in \cite{AD}. Moreover,
it has been shown \cite{AD} that $P_a$ and $Q_a$ are compatible through
the method of prolongation  and describe the hierarchy equations (\ref{laxeq}) as follows
\be
\pa_{t_n}
\left(
\ba{c}
\Phi_a\\
\Psi_a
\ea
\right)=
P_a
\left(
\ba{c}
\frac{\de H_{n+1}}{\de \Phi_a}\\
\frac{\de H_{n+1}}{\de \Psi_a}
\ea
\right)=
Q_a
\left(
\ba{c}
\frac{\de H_{n}}{\de \Phi_a}\\
\frac{\de H_{n}}{\de \Psi_a}
\ea
\right)
\ee
where the Hamiltonian $H_n$ are defined by $H_n=-\frac{1}{n}\str L_a^n$.
Hence, the gauge transformation $H_a$ (or $G_a$) is a canonical map.

Next, let us turn to the gauge transformation $H_b$. From (\ref{hb}), the
linearized map $H_b'$ and its transposed map $H_b'^{\dag}$ can
be constructed as follows
\be
H_b'=
\left(
\ba{cc}
-\Phi_b D^{-1}-\Psi_b^{-1}\pa & \Psi_b^{-1}\\
\Psi_b D^{-1} & 0
\ea
\right),\qquad
H_b'^{\dag}=
\left(
\ba{cc}
\pa\Psi_b^{-1}+D^{-1}\Phi_b & -D^{-1}\Psi_b\\
\Psi_b^{-1} & 0
\ea
\right)
\label{linhb}
\ee
where $\Phi_b$ and $\Psi_b$ are related to $J_0$ and $J_1$ via 
(\ref{dicgb}) (or (\ref{dichb})). Using (\ref{linhb}),
we can obtain two Poisson structures of the sAKNS hierarchy for the
case (b). After some algebras, we have
\bea
H_b'\Th_1 H_b'^{\dag} &=&
\left(
\ba{cc}
0 &1\\
-1 & 0
\ea
\right)\equiv -P_b 
\label{canpb}\\
H_b'\Th_2 H_b'^{\dag} &=&
\left(
\ba{cc}
-\Phi_b D^{-2}(D\Phi_b)-(D\Phi_b)D^{-2}\Phi_b & 
D^2+\Phi_b D^{-2}\Psi_b D+(D\Phi_b)D^{-2}\Psi_b  \\
-2\Phi_b D^{-2}\Phi_b\Psi_b D^{-2}\Phi_b      & 
+2\Phi_b D^{-2}\Phi_b\Psi_b D^{-2}\Psi_b                 \\
\mbox{} & \mbox{} \\
D^2+D\Psi_b D^{-2}\Phi_b+\Psi_b D^{-2}(D\Phi_b) &  
-\Psi_b D^{-1}\Psi_b D-D\Psi_b D^{-2}\Psi_b  \\
+2\Psi_b D^{-2}\Phi_b\Psi_b D^{-2}\Phi_b     &  
-2\Psi_b D^{-2}\Phi_b\Psi_b D^{-2}\Psi_b
\ea
\right)\equiv -Q_b
\label{canqb}
\eea
which imply that the hierarchy equations (\ref{laxeq}) 
for case (b) can be written as
\be
\pa_{t_n}
\left(
\ba{c}
\Phi_b\\
\Psi_b
\ea
\right)=
P_b
\left(
\ba{c}
\frac{\de H_{n+1}}{\de \Phi_b}\\
\frac{\de H_{n+1}}{\de \Psi_b}
\ea
\right)=
Q_b
\left(
\ba{c}
\frac{\de H_{n}}{\de \Phi_b}\\
\frac{\de H_{n}}{\de \Psi_b}
\ea
\right).
\label{hameqb}
\ee
Note that the minus sign appearing in the front of $P_b$ and $Q_b$ in
(\ref{canpb}) and (\ref{canqb}) is due to the fact that 
the parity of the gauge operator of the 
gauge transformation $H_b$ is odd. Therefore, 
by the identity $\str AB=(-1)^{|A||B|}\str BA$, the Hamiltonians
$H_n$ in (\ref{hameqb}) are equal to the minus ones for the sTB hierarchy.
We follow the same line in \cite{AR} to investigate the Jacobi-identity
for $P_b$ and $Q_b$ by using the prolongation method.
It turns out that $P_b$ and $Q_b$ are compatible and indeed
define a bi-Hamiltonian structure of the associated hierarchy.
Hence, just like $H_a$, the gauge transformation $H_b$ is 
canonical as well. 

To sum up, the canonical property of the gauge transformations between
the sAKNS and sTB hierarchies can be summarized as follows
\be
H_i'\Th_1 H_i'^{\dag}=(-1)^{|H_i|}P_i, \qquad
H_i'\Th_2 H_i'^{\dag}=(-1)^{|H_i|}Q_i, \qquad i=a,b.
\label{equiv}
\ee 

\section{Darboux-B\"acklund transformations}
Having constructed the canonical gauge transformations between the sAKNS 
and sTB hierarchies, now we would like to use  these gauge transformations
to derive the DBTs for the sAKNS hierarchy  itself.

Given a sAKNS Lax operator, say $L_a$, we can perform the gauge
transformation $G_a$ followed by $H_b$ to obtain the Lax operator $L_b$
as follows
\be
L_a \stackrel{G_a}{\rightarrow} K \stackrel{H_b}{\rightarrow} L_b
\ee
That is, using (\ref{ga}) and (\ref{hb}), we can define the gauge operator 
$T(\Phi_a)=\Phi_a D\Phi_a^{-1}$ such that
\bea
L_a\rightarrow L_b &=&TL_a T^{-1}\\
&\equiv&\pa+\Phi_b D^{-1}\Psi_b\no
\eea
where the (adjoint) eigenfunctions are related by
\bea
\Phi_b &=&\Phi_a(\Phi_a\Psi_a+(D^3\ln\Phi_a))
\label{sab}\\
\Psi_b&=&\Phi_a^{-1}
\eea
Notice that although the gauge transformation (\ref{sab}) preserves the form
of the Lax operator and the Lax  formulations, however, the parity of the 
transformed (adjoint) eigenfunction has been changed due to the
fact that the parity of the  gauge operator $S_{ab}$ is odd. 
Thus, strictly speaking, the gauge transformation (\ref{sab}) is not a DBT but
``quasi-DBT".

On the other hand, we can construct another quasi-DBT from $L_b$ to $L_a$ as follows
 \be
L_b \stackrel{G_b}{\rightarrow} K \stackrel{H_a}{\rightarrow} L_a
\ee
which  is triggered by the gauge operator 
$S(\Psi_b)=\Psi_b^{-1}D^{-1}\Psi_b$ such that
\bea
L_b\rightarrow L_a &=&SL_b S^{-1}
\label{sba}\\
&\equiv& \pa+\Phi_a D^{-1}\Psi_a\no
\eea
where
\bea
\Phi_a&=&\Psi_b^{-1}\\
\Psi_a&=&\Phi_b(\Phi_b\Psi_b+(D^3\ln\Psi_b)).
\eea
Note that both quasi-DBTs (\ref{sab}) and (\ref{sba}) are canonical
since they are constructed out from the canonical transformations
$G_i$ and $H_i$. We also remark that the form of the gauge operator $T$
was first considered in \cite{Liu} for studying the DBT for the Manin-Radul
super KdV equation\cite{MR}. 

Motivated by the above discussions, we may have true DBTs by considering  the 
hierarchy equations (\ref{laxeq}) associated with the Lax operator
\be
L=\pa+\Phi_1 D^{-1}\Psi_1+\Phi_2 D^{-1}\Psi_2
\ee
with parity $|\Phi_1|=|\Psi_2|=0$ and $|\Psi_1|=|\Phi_2|=1$. 
Let us consider the DBT triggered by the eigenfunction $\Phi_1$ as follows
\bea
L\rightarrow \hat{L}&=&TLT^{-1},\qquad T(\Phi_1)\equiv\Phi_1 D\Phi_1^{-1}\\
&\equiv& \pa+\hat{\Phi}_1 D^{-1}\hat{\Psi}_1+\hat{\Phi}_2 D^{-1}\hat{\Psi}_2\no
\eea
where the transformed (adjoint) eigenfunctions  are given by
\bea
{\hat{\Phi}_1}&=&\Phi_1(D\Phi_1^{-1}\Phi_2)=(T(\Phi_1)\Phi_2)\no\\
{\hat{\Psi}_1}&=&\Phi_1^{-1}(D^{-1}\Phi_1\Psi_2)=(S(\Phi_1)\Psi_2)\no\\
{\hat{\Phi}_2}&=&\Phi_1(\Phi_1\Psi_1-\Phi_2\Psi_2+(D^3\ln\Phi_1)
+(D\Phi_1^{-1}\Phi_2)(D^{-1}\Phi_1\Psi_2))=(T(\Phi_1)L\Phi_1)\no\\
{\hat{\Psi}_2}&=&\Phi_1^{-1}
\eea
with parity $|\hat{\Phi}_1|=|\hat{\Psi}_2|=0$ and $|\hat{\Psi}_1|=|\hat{\Phi}_2|=1$. 

Moreover, we can consider the DBT triggered by the adjoint eigenfunction $\Psi_2$
as follows
\bea
L\rightarrow \hat{L}&=&SLS^{-1},\qquad S(\Psi_2)\equiv\Psi_2^{-1} D^{-1}\Psi_2\\
&\equiv& \pa+\hat{\Phi}_1 D^{-1}\hat{\Psi}_1+\hat{\Phi}_2 D^{-1}\hat{\Psi}_2\no
\eea
where
\bea
{\hat{\Phi}_1}&=&\Psi_2^{-1}\no\\
{\hat{\Psi}_1}&=&(\Phi_2\Psi_2-\Phi_1\Psi_1+(D^3\ln\Psi_2)+
(D^{-1}\Psi_2\Phi_1)(D\Psi_1\Psi_2^{-1}))\Psi_2=-(T(\Psi_2)L^*\Psi_2)\no\\
{\hat{\Phi}_2}&=&\Psi_2^{-1}(D^{-1}\Psi_2\Phi_1)=(S(\Psi_2)\Phi_1)\no\\
{\hat{\Psi}_2}&=&\Psi_2(D\Psi_2^{-1}\Psi_1)=(T(\Psi_2)\Psi_1)
\eea
with parity $|\hat{\Phi}_1|=|\hat{\Psi}_2|=0$ and $|\hat{\Psi}_1|=|\hat{\Phi}_2|=1$. 

Finally, we would like to mention that the above scheme can be generalized to 
a class of supersymmetric hierarchies which have Lax operators of the form
\be
L=\pa+\sum_{i=1}^n(\Phi_{2i-1}D^{-1}\Psi_{2i-1}+\Phi_{2i}D^{-1}\Psi_{2i}),\qquad
(n\geq 1)
\ee
with parity  $|\Phi_{2i-1}|=|\Psi_{2i}|=0$ and $|\Phi_{2i}|=|\Psi_{2i-1}|=1$.
The gauge operators of the DBTs then can be constructed from the even
(adjoint) eigenfunctions as $T_i=\Phi_{2i-1}D\Phi_{2i-1}^{-1}$ 
or $S_i=\Psi_{2i}^{-1}D^{-1}\Psi_{2i}$ which not only preserve the Lax
formulations but also the parity content of the (adjoint) eigenfunctions
in the Lax operator. 

\section{Concluding remarks}

We have established the gauge equivalences between the sAKNS and sTB hierarchies.
We have also shown that the gauge transformations connecting these two hierarchies
are canonical, in the sense that the bi-Hamiltonian structure of the sAKNS hierarchy
is mapped to the bi-Hamiltonian structure of the sTB hierarchy according to 
Eq. (\ref{equiv}). Using these gauge transformations, the (quasi) DBTs for
the sAKNS hierarchy and its generalizations  can be constructed, which 
 turns out to be canonical as well. Some other topics such as iterated DBTs
,soliton solutions and nonlocal conserved charges of these hierarchies are 
worth further investigation\cite{ST}. We leave this work to a future publication.

{\bf Acknowledgments\/}
We would like to thank Prof. W.-J. Huang for helpful discussions.
This work is supported by the National Science Council of Taiwan 
under grant No. NSC-87-2811-M-007-0025.
M.H. Tu also wish to thank Center for Theoretical Sciences of National
Science Council of Taiwan for partial support.

\end{document}